\documentclass[conference]{IEEEtran}

\usepackage[utf8]{inputenc}
\usepackage{amsmath,amssymb,amsfonts}
\usepackage{algorithmic}
\usepackage{graphicx}
\usepackage[export]{adjustbox}
\usepackage{textcomp}
\usepackage{xcolor}
\definecolor{refC}{rgb}{0,0,0.75}
\usepackage[unicode,colorlinks,linkcolor=refC]{hyperref}
\usepackage{dblfloatfix}

\usepackage{booktabs}
\usepackage{multirow}

\newcommand*{\fullref}[1]{\hyperref[{#1}]{\autoref*{#1}: \nameref*{#1}}}
\newcommand*{\shortref}[1]{\hyperref[{#1}]{\nameref*{#1} \ref*{#1}}}

\def\BibTeX{{\rm B\kern-.05em{\sc i\kern-.025em b}\kern-.08em
    T\kern-.1667em\lower.7ex\hbox{E}\kern-.125emX}}

\IEEEoverridecommandlockouts\IEEEpubid{\makebox[\columnwidth]{} \hspace{\columnsep}\makebox[\columnwidth]{ }}

\begin{document}

\title{Boundaries of Flow Table Usage Reduction Algorithms Based on Elephant Flow Detection}

\author{
\IEEEauthorblockN{Piotr Jurkiewicz}
\IEEEauthorblockA{\textit{Department of Telecommunications} \\
\textit{AGH University of Science and Technology}\\
Kraków, Poland \\
piotr.jurkiewicz@agh.edu.pl}
}

\maketitle

\begin{abstract}

The majority of Internet traffic is caused by a relatively small number of flows (so-called elephant flows). This phenomenon can be exploited to facilitate traffic engineering: resource-costly individual flow forwarding entries can be created only for elephants while serving mice over the shortest paths. Although this idea already appeared in proposed TE systems, it was not examined by itself. It remains unknown what extent of flow table occupancy and operations number reduction can be achieved or how to select thresholds or sampling rates to cover the desired fraction of traffic. In this paper, we use reproducible traffic models obtained from a 30-day-long campus trace covering 4 billion flows, to answer these questions. We establish theoretical boundaries for flow table usage reduction algorithms that classify flows since the first packet, after reaching a predefined counter threshold or detect elephants by sampling. An important finding is that simple packet sampling performs surprisingly well on realistic traffic, reducing the number of flow entries by a factor up to 400, still covering 80\% of the traffic. We also provide an open-source software package allowing the replication of our experiments or the performing of similar evaluations for other algorithms or flow distributions.

\end{abstract}


\section{Introduction}

It is widely believed that the distribution of flow lengths and sizes in the Internet follows the Pareto principle: the majority of traffic is comprised of a relatively small number of flows. Such flows are called \emph{elephant flows}. The remaining flows, which are large in number but carry very little traffic, are called \emph{mice flows}. In practice, flow length and size distributions are even more long-tailed than the Pareto rule (80/20) assumes. According to recent analysis, 80\% of traffic is caused by only 0.2-0.4\% of flows \cite{megyesi2013analysis}\cite{flows-agh}.


This phenomenon can be exploited to facilitate quality of service (QoS) provisioning and traffic engineering (TE). In per-packet routing, due to loop prevention constraints, only a subset of disjoint paths existing between selected nodes can be used \cite{dual}. Adaptive (load-sensitive) routing is also impossible in a per-packet approach, as the dynamic alteration of link costs leads to instability (route flapping) which ultimately deteriorates network performance; this has been shown by early ARPANET pitfalls \cite{Bertsekas82} and definitively proven in \cite{wang1992analysis}.

Flow-based routing can overcome these problems by maintaining separate per-flow forwarding entries. It allows flows between the same endpoints to follow any number of alternative paths. Adaptive flow routing is also more stable than selecting paths at the packet level, as the load on each link fluctuates more slowly, as has been shown in \cite{FAMTAR_IMPLEMENTATION}. To sum up, with flow-based routing more traffic can be served using existing infrastructure, reducing the need for links oversubscription.

However, despite recent technological advancements, the number of simultaneous flows in networks still overwhelms the capacities of switch flow tables \cite{shen2019powerful}. Moreover, in the case of centralized control plane usage, controller throughput can impose additional limits on the rate of incoming flows. This is confirmed in practice. The real deployments, like Google's B4 \cite{jain2013b4}, are limited to proactive systems. Such systems forward packets according to predefined, per-prefix shortest-path entries. Specific entries are created for heavy-hitter aggregates, which are detected basing on external information (for example, a notification of an expected migration between datacentres). In the case of non-private WANs, such a piece of information is not available. A possible solution would be to focus on dynamically identified elephant flows. This would significantly reduce the number of flow entries while simultaneously keeping most traffic covered by TE.



The problem is \emph{early} identification of whether a particular flow is, or more precisely will become an elephant. Issues related to elephant flow detection have been the subject of many papers. However, most works have been directed in the context of flow accounting and monitoring and focus on detection accuracy. In the case of traffic engineering, detection accuracy is not the most important issue. Instead, the focus should be put on the moment of identification, and particularly on the amount of traffic transmitted by the flows after their classification as elephants (i.e. when they have individual entries) and the resulting reduction in flow table occupancy. None of the previous works analyzes these parameters or provides numerical boundaries for them.

Moreover, any elephant-related mechanism performance strictly depends upon flow length and size distribution. Papers make different, and often over-simplistic and arbitrary assumptions, for example by considering constant elephant-to-mice ratio and sizes, which do not correspond with reality. Other papers use distributions obtained from real traffic traces; however, they are either irreproducible or simplified to a single distribution function.

The definition of elephant flow is another factor, which differs among many papers. The analysis presented in this paper is independent of the elephant flow definition. We did not assume a fixed elephant definition, because we do not focus on strict classification of flows as mice or elephants. Instead, our goal is the statistical reduction of flow table occupancy.

This paper aims to fill the gap in research and provide numerical boundaries for the performance of flow table usage reduction algorithms. The key point of our research is the use of realistic, accurate, and reproducible flow length and size distributions. We analyze the performance of three approaches, which can provide boundaries to several classes of elephant detection algorithms:

\begin{itemize}

\item \emph{first}, which assumes a pre-established knowledge about flow length/size and classifies it accordingly since its first packet,

\item \emph{threshold}, which classifies a flow as an elephant after transmitting a predefined amount of packets or bytes,

\item \emph{sampling}, which performs packet sampling and classifies flows in a probabilistic manner.

\end{itemize}

\smallskip

Similar ideas have appeared already as a part of proposed TE systems (see \fullref{related-works}). However, metrics in these papers were related to overall TE system performance (like packet loss). The trade-off between table occupancy reduction and traffic coverage has not been analyzed. Therefore, the novelty of this paper lies in:

\begin{itemize}

\item Analysis of parameters relevant to traffic engineering in the context of SDN: the fraction of traffic covered, reduction of the number of flow entry operations (and thus controller traffic), and reduction of flow table occupancy.

\item Providing theoretical upper and lower boundaries for several classes of flow table usage reduction algorithms based on elephant flow classification.

\item Use of realistic and accurate distribution mixtures obtained from 30-day-long 4-billion flow trace of campus/residential traffic, which is many orders of magnitude more than in previous research.

\item Reproducibility of the research, as both the distribution mixtures and code used for analysis are provided as an open-source package \cite{github-flow-models}.

\end{itemize}

\section{Related works}\label{related-works}

The idea of performing adaptive routing only for elephants while keeping mice on the shortest paths is not new. According to our knowledge, it was first proposed in 1999 in \cite{rexford-long-flows}. The authors, however, did not solve the problem of detecting elephant flows. They propose the usage of per-flow counters or timers, which is pointless considering that our goal is the reduction in the number of tracked flows and flow table operations. Moreover, their analysis is based on a one-week trace collected in 1997, which is both outdated and too short. A similar approach was proposed in \cite{sarrar2012leveraging}, but concerned the top destination IP prefixes (so-called \emph{heavy hitters}) instead of 5-tuple flows.


This approach has been recently reiterated, specifically in the SDN context. The general idea is to initially install shortest path wildcard entries and monitor the traffic in order to identify elephant flows. After identification, the controller can compute alternative non-congested paths for them based on the global network view and install individual entries for these most significant flows in order to load-balance traffic.

Hedera \cite{al2010hedera} was proposed as a dynamic flow scheduling system for datacentres, aimed at going beyond equal-cost multipath (ECMP) routing limitations. By default, all flows are load-balanced onto ECMP paths. Such a path is used until the flow grows and meets a predefined threshold rate. After reaching the threshold, elephants are rerouted in mid-connection onto flow-specific paths, computed dynamically by the controller. Hedera assumes that the edge switches collect flow statistics for all flows using OpenFlow counters. This means that it actually only reduces non-edge switches overhead, while the edge switches still have to maintain individual entries for all flows.


In 2011, Curtis et al. presented a system called Mahout \cite{curtis2011mahout}. Unlike Hedera, it performs elephant detection at the end hosts by monitoring socket buffers (via a shim layer in the OS). After reaching a predefined threshold, it marks subsequent packets of flow using an in-band signaling mechanism. The switches in the network are configured to forward these marked packets to the controller, which as with Hedera computes the best path and installs flow-specific entries in switches. With that approach, overhead can be eliminated from the switches and controller, as elephant detection is moved to end hosts, which however must be modified.



DevoFlow \cite{curtis2011devoflow} is another example of a complete TE system based on modified Openflow switches, which key feature is focusing on significant flows. To detect these flows, it explores both \emph{threshold} and \emph{sampling} approaches, which is similar to this paper. However, the traffic model used in DevoFlow was a datacentre workload, which considerably differs from the residential/ISP load. Moreover, the authors "reverse engineered" the flow distributions they used from plots presented in another paper and did not make them available, which makes their results both inaccurate and unreproducible. Authors do not analyze the amount of traffic covered. Instead, the only performance indicator provided is the aggregate throughput of the whole network, which also depends on topology, demands matrix, and routing decisions. Only absolute values of the number of flow entries are provided, so the relative reduction of table occupancy also cannot be determined. Only three thresholds/sampling probabilities are analyzed, whereas our paper provides analysis for the continuous spectrum of values.

A similar system to DevoFlow is proposed in \cite{xu2017scalable}. It detects elephant flows only on edge switches with the use of a Bloom filters variant, called \emph{randomized counter sharing}. However, the traffic model used for the evaluation of this mechanism is oversimplified: the authors assume the power law for the flow-size distribution, where 20\% of all flows account for 80\% of traffic volume. This is, as shown in \cite{megyesi2013analysis} and \cite{flows-agh}, far from reality.


OpenSample \cite{suh2014opensample} is a TE system based on the rerouting of elephant flows, which are detected by sampling packet headers on switches with sFlow. The authors claim that it can achieve a low latency of measurements with a high degree of accuracy. Unlike Mahout and MicroTE, OpenSample can be implemented without end host modifications and unlike Hedera, it does not require the use of expensive OpenFlow counters. The authors provide an analysis of the percentage of traffic covered after detection and after rerouting. However, the used traffic model is extremely simplified. They consider only two classes of flows, short flows with an exponential distribution of mean 1 MB size, and long flows with the same exponential distribution but a 1 GB mean flow size. The paper also does not analyze flow table occupancy reduction.


Planck \cite{rasley2014planck} is another system, similar to OpenSample. It deserves special attention because it does not use sFlow for packet sampling. Instead, the authors propose the use of a port mirroring feature to redirect all packets to a single monitoring port. Because the total traffic forwarded through the switch usually exceeds the capacity of the monitoring port, some packets are dropped, which effectively provides a packet sample. Such an approach has several advantages over sFlow, specifically, the reduced load of switch CPU and significantly lower latencies. As a result of this, the Planck-based traffic engineering system can reroute congested flows within milliseconds, which can improve its overall performance.

The following surveys provide a good overview of SDN traffic engineering systems: \cite{akyildiz2014roadmap} \cite{abbasi2016traffic}. The use of packet sampling for traffic engineering purposes is indicated in \cite{cern-sampling-report}, which also provides a good overview of other packet sampling techniques. It has to be noted that several other, non-sampling based techniques for elephant/heavy-hitters detection and flow table occupancy reduction have been proposed. This includes flow table compression and entry aggregation \cite{liu2010tcam} \cite{7810727} \cite{dulinski2020mpls}, entry caching \cite{katta2016cacheflow}, label-based switching \cite{huang2018software}, use of multiple hash tables \cite{sivaraman2017heavy} \cite{wang2019distributed} \cite{da2018ideafix} and a variety of Bloom filters or sketching-based approaches \cite{CAFARO2019770} \cite{huang2017sketchvisor} \cite{basat2017optimal} \cite{10.1145/3359989.3365408}.







Moreover, in the wake of growing machine learning popularity, it recently started being employed as well. Going specifically to flow classification based on the first packet, in \cite{9016231} authors propose a machine learning based flow classification based on for the NFV offloading. Classification is performed using features extracted from the header of the first packet (5-tuple, size of the first packet). In \cite{da2019predicting} prediction of new flows size and duration is done at the flow start through a Locally Weighted Regression (LWR) model, using the previous flows behavior and its temporal correlation with the new flow.

%
%

To sum up, most of the works mentioned in previous paragraphs focus on the issue from the network monitoring or accounting point of view. The main evaluated parameters are related to detection accuracy; it includes false positive ratio, false-negative ratio, and the precision of flow length/size estimation from the sample. On the other hand, papers focusing on the concept from SDN and traffic engineering point of view, examine the performance of the TE whole system and analyze metrics like network throughput or packet loss. Therefore, despite the idea already appeared as a part of proposed systems, it has not been examined in isolation. Its numerical performance boundaries are yet to be explored.

\begin{figure*}[!ht]
\minipage{0.33\textwidth}
  \vspace{0.85cm}
  \includegraphics[width=\linewidth]{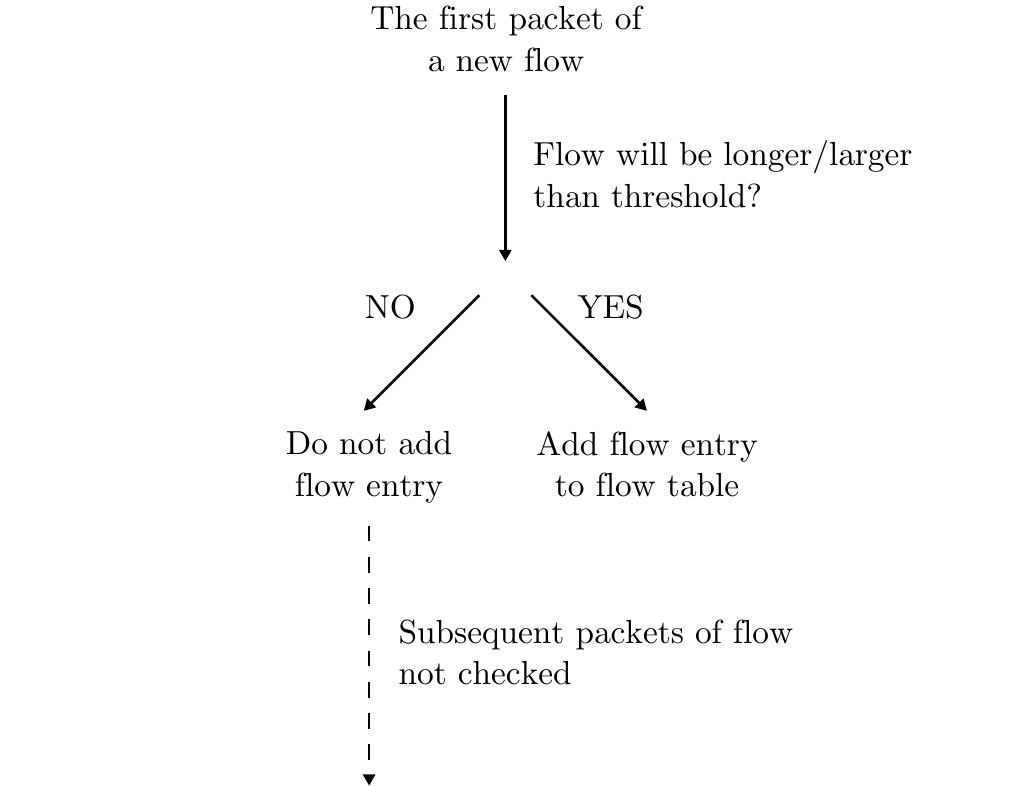}
  \caption{The \emph{first} algorithm.}
  \label{schema-first}
\endminipage\hfill
\minipage{0.33\textwidth}
  \includegraphics[width=\linewidth]{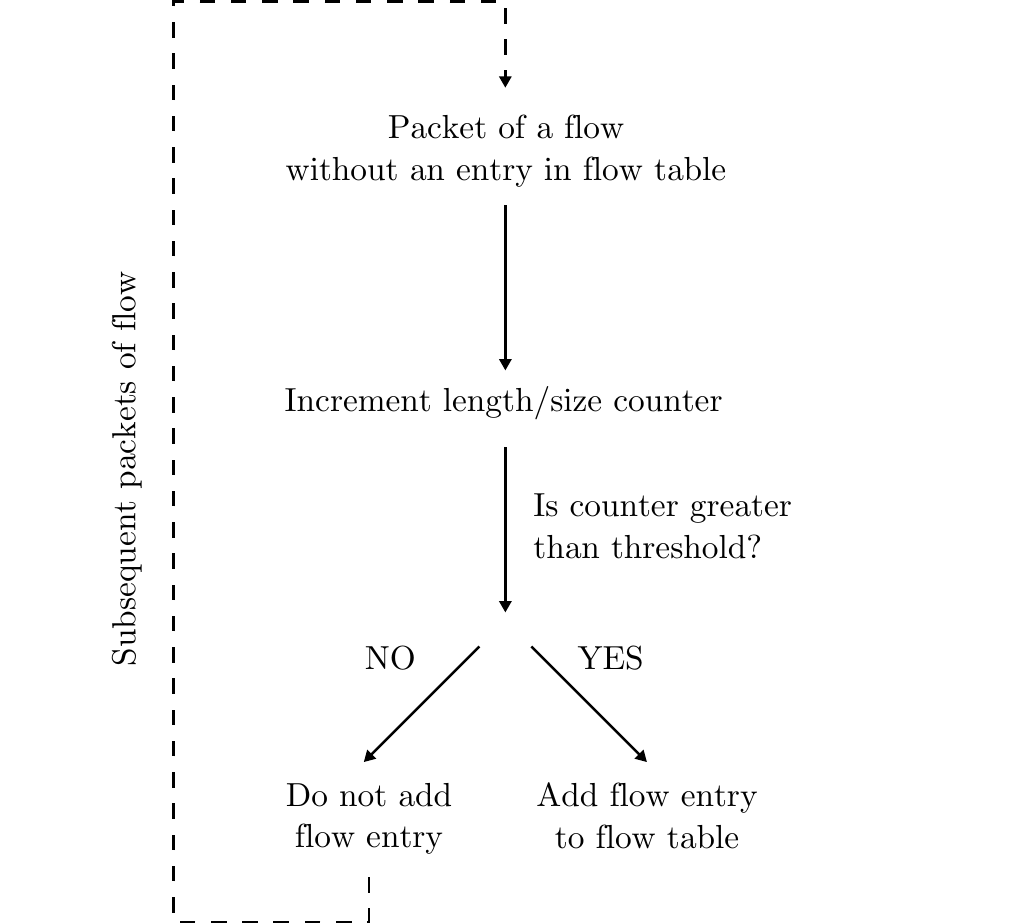}
  \caption{The \emph{threshold} algorithm.}
  \label{schema-threshold}
\endminipage\hfill
\minipage{0.33\textwidth}%
  \includegraphics[width=\linewidth]{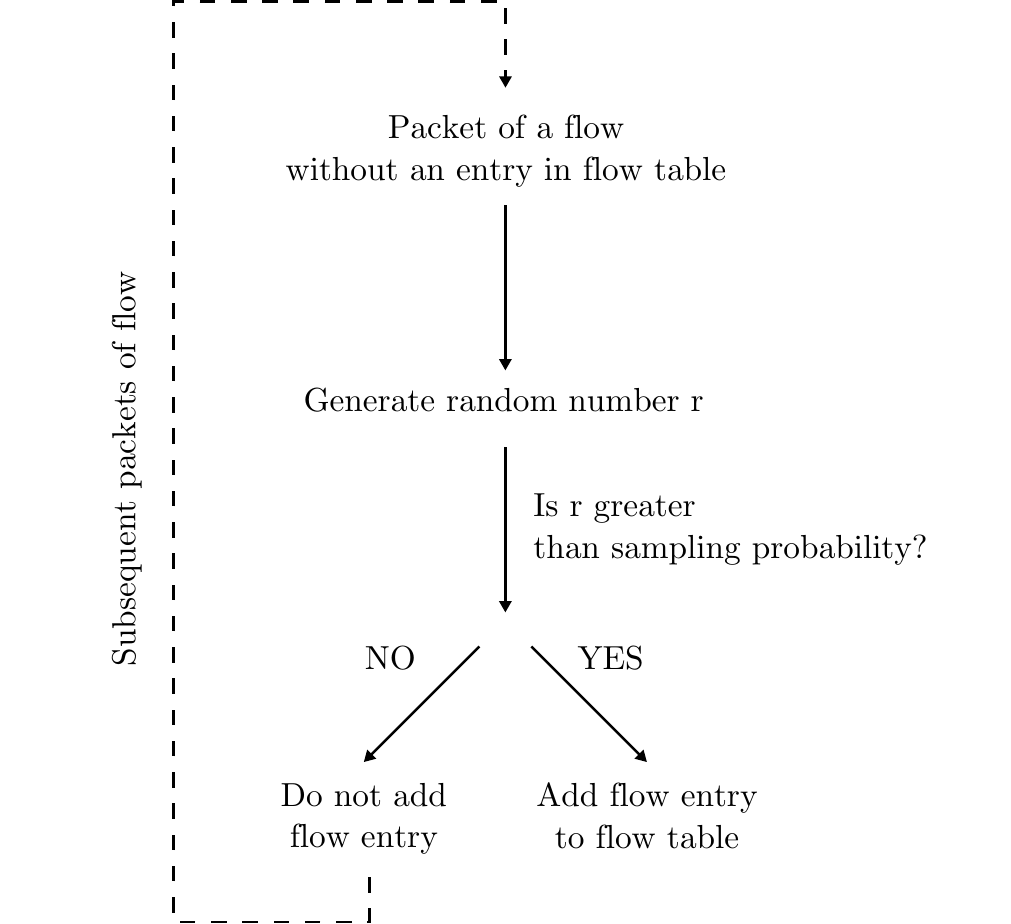}
  \caption{The \emph{sampling} algorithm.}
  \label{schema-sampling}
\endminipage
\end{figure*}

\section{Analyzed algorithms}
\label{algorithms}

\subsection{First}

In the \emph{first} algorithm, a flow entry is created on the arrival of the flow's first packet when the length/size of flow will exceed the selected threshold. Thus, it assumes that based on the first packet, the length/size of the whole flow can be determined. The flowchart of the \emph{first} algorithm is show in \autoref{schema-first}.

It is impossible to achieve in reality: it is not possible to know in advance how long a flow will be, despite that classification with the first packet would be the most beneficial. Classifying a flow correctly with the first packet would allow avoiding rerouting it in the middle. Flow rerouting may result in transport layer issues. Specifically, packets may become reordered, which can cause their retransmission. Additionally, path change can misdirect congestion control algorithms, which usually work basing on observed path delay and throughput estimation.

However, the \emph{first} approach is still worth evaluating, because it can provide an upper boundary for the performance of all elephant detection algorithms. Moreover, an (imperfect) flow classification with the first packet can be achieved with the help of machine learning based algorithms, which attempt to predict the flow length/size basing on header values. Such algorithms are becoming a hot research topic, especially in combination with their realization in programmable dataplanes, like P4 \cite{da2018ideafix}\cite{hardegen2020ieee}. Thus, the results of \emph{first} approach will provide an upper baseline for all of these research.

\subsection{Threshold}

As already mentioned, in practice switches cannot know in advance whether a newly appeared flow will eventually become an elephant or a mice flow. Thus, they cannot create an entry for this flow when its first packet appears. Instead, flow entry can be created when the amount of traffic or the number of transmitted packets exceeds a certain elephant detection threshold. The most trivial approach is to use per-flow counters on each switch. A counter reaching a threshold would cause a flow entry to be created. This is the outline of \emph{threshold} algorithm, which is shown in \autoref{schema-threshold}.

Such an algorithm is realizable, however, it has obvious drawbacks. Per-flow counters must be stored and updated with each packet. In the case of OpenFlow, flow entries aimed at packet counting use the same memory as any other flow entry, so it would not yield any reductions in flow table usage. In the case of other data plane technologies, the low-overhead implementation of accurate per-flow counters may be possible, as shown by TurboFlow for P4 \cite{turboflow}. Moreover, solutions based on inexact counters, like Bloom filters or sketches, can also classify elephant flows after reaching a predefined threshold, at the same time requiring a reduced amount of memory. Thus, the results of the \emph{threshold} algorithm can provide an upper bound for all algorithms, which are based of any kind of counting. This includes exact counters and inexact counting methods, based on Bloom filters or sketching.


\subsection{Sampling}

An alternative approach is to use sampling. Packets without entries in flow tables can be randomly sampled with some probability $ p $. If a packet is sampled, a new flow entry is created and subsequent packets of that flow are forwarded in accordance with it, without being sampled. Otherwise, the packet is forwarded basing on aggregated (usually ECMP) entry without the creation of a flow entry, and sampling is performed for the rest of the packets until a flow entry is created, as shown in \autoref{schema-sampling}.

The probability that a flow has an entry in the flow table after reaching $ n $ packets is given by:

\[
p_{total} = 1 - (1 - p)^n
\]

This means, that by adjusting $p$, a network operator can adaptively control how many flow entries will be stored in tables and how much traffic will be covered by them.

The \emph{sampling} algorithm can also be used to sample flows according to their size. In this way, larger packets have a greater probability of being sampled. Such an approach is called Non-Uniform Probabilistic Sampling in RFC 5475 \cite{rfc5475}. It is enough to scale the sampling probability for each packet proportionally to its size. In our simulation, we scaled provided sampling probabilities by relative packet size:

\[
p_{scaled} = p \cdot \frac{s}{s_{max}}
\]

where:

\begin{itemize}

\item $s$ is the packet size, and
\item $s_{max}$ is the maximum packet size.

\end{itemize}

In our calculations and simulations, we assume that sampling is performed only on edge switches (i.e. any given packet can be sampled only once when it enters the network). An alternative setup would be a network in which all switches sample packets independently (i.e. a packet is sampled on all switches on its path). In such a case, the effective sampling probability value can be calculated using the following equation:

\[
p_{eff} = \sum_{k \in paths} P^{(k)} \cdot (1 - \prod_{i=1}^{l^{(k)}} (1-p_i^{(k)}))
\]

where:

\begin{itemize}

\item $p_{eff}$ is the effective sampling probability,
\item $P^{(k)}$ is the probability of going through path $k$,
\item $p_i^{(k)}$ is the probability of being sampled at switch $i$ of path $k$, and
\item $l^{(k)}$ is the length of this path.

\end{itemize}

Assuming that the probability of going through the each path is the same (i.e. each switch is traversed by a high number of flows, going through a variety of paths), the above equation can be simplified to:

\[
p_{eff} = 1 - (1 - p)^{l_{avg}}
\]

where:

\begin{itemize}

\item $p$ is the sampling probability set by the operator on a single switch, and
\item $l_{avg}$ is the average length of path in the network.

\end{itemize}

The $p_{eff}$ value, calculated with the above equation, can be used to read values from tables presented in \shortref{appendix} to determine what traffic coverage and occupancy reduction will be achieved with selected $p$ in the case of all-switch sampling.

The advantage of \emph{sampling} method is that it is stateless, i.e. there is no need to store and update any kind of counters. Additionally, it has a negligible performance impact, as random number generation can be performed using a hardware random generator or a simple software pseudorandom generator with a few CPU cycles. The P4 standard allows to perform sampling in dataplane\footnote{\url{https://p4.org/p4-spec/docs/PSA.html\#sec-random}}. Alternatively to on-switch sampling, a low-overhead port mirroring based approach can be used, as proposed in Planck system \cite{rasley2014planck}. However, as classification happens not with the first packet, but in the middle of a flow, it may cause rerouting resulting in transport layer issues. Nevertheless, it provides a lower boundary for all algorithms which classify flows not on the first packet.

\section{Results}
\label{results}

In this section, we present the results of the simulations, as well as analytical calculations. The following parameters are analyzed:

\begin{itemize}

\item \emph{traffic coverage} -- the percentage of traffic (bytes) in the network which were transmitted by flows after their detection (i.e. when they had an individual entry),

\item \emph{operations reduction} -- the factor by which the number of flow entry additions/removals (and thus the controller traffic) can be reduced,

\item \emph{occupancy reduction} -- the factor by which the average number of flow entries in tables can be reduced.

\end{itemize}

All these values are presented relative to the baseline case, in which all flows have their entries created with their first packets (the classic SDN reactive mode). In our analysis we assume that all flows have the same rate. This means that, as long we focus only on relative performance metrics, any flow aging issues are irrelevant here.

As already mentioned, flow size and length distributions have a crucial impact on the performance of any elephant-related algorithms. In this paper we use flow length and size distributions from the \texttt{agh\_2015} dataset presented in \cite{flows-agh}. These are based on traffic traces collected on the outgoing interface of the campus/residential network. Unlike CAIDA traces, which are truncated to one hour, which distorts tails of distributions and therefore makes them unsuitable for such analysis, they were collected over a continuous period of 30 days and consisting of more than 4 billion flows. Both the timespan of the collection and the number of flows are many orders of magnitude higher than in previously published flow statistics. These distributions are in line with selected values of CAIDA and BME traces presented in \cite{megyesi2013analysis}, which confirms their credibility. The authors of \cite{flows-agh} provide accurately fitted distribution mixture models. They allow an analytical calculation of all the performance parameters of the evaluated algorithms, which also would be impossible with the CAIDA data. We also use them to generate flow samples in packet-level simulations.

For selected values of thresholds (in cases of \emph{first} and \emph{threshold} algorithms) and sampling probability (in the case of \emph{sampling} algorithm) we performed packet-level simulations. Packets were randomly generated, basing on the distribution mixtures from the used traffic model. For each value of threshold/sampling probability, we performed the experiment five times with different random seeds, each time generating 1 billion flows, and calculated mean values from these five runs. All algorithms are evaluated using both flow length (number of packets) and flow size (amount of bytes) as an elephant classification criterion. Results of these simulations are shown in \autoref{tab-length} and \autoref{tab-size} provided in \shortref{appendix}.

\begin{figure*}[!htb]
\centering
\includegraphics[scale=0.46,right]{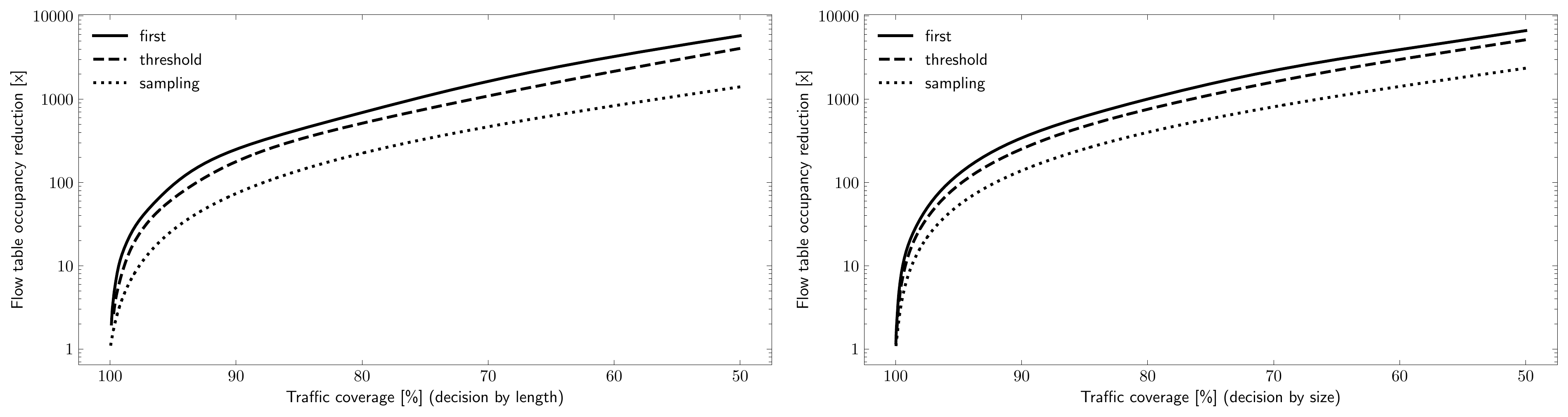}
\caption{Analytically calculated flow table occupancy reduction in function of traffic coverage.}
\label{fig-occupancy-absolute}
\end{figure*}

\begin{figure*}[!htb]
\centering
\includegraphics[scale=0.46,right]{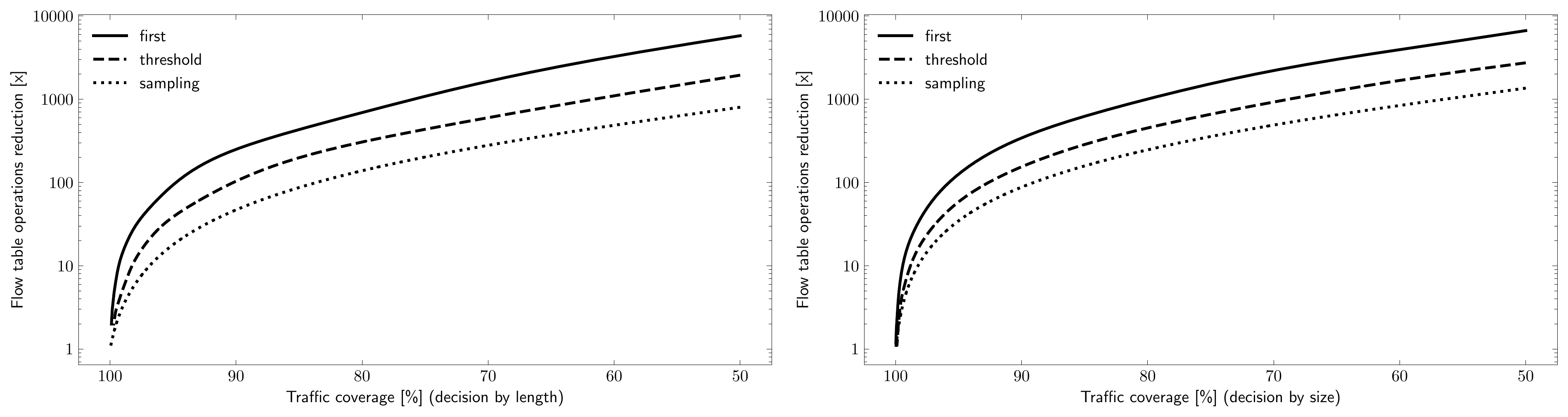}
\caption{Analytically calculated flow table operations number reduction in function of traffic coverage.}
\label{fig-operations-absolute}
\end{figure*}

Additionally, we use distribution mixture equations provided in \cite{flows-agh} to calculate the performance for a continuous spectrum of threshold/sampling probability values, which is impossible with simulations. This allows straightforward plotting and comparison of algorithms performance against each other. Analytically calculated reduction of flow table occupancy is shown in \autoref{fig-occupancy-absolute} on y-axis (logarithmic). Similarly, reduction in number of operations is presented on \autoref{fig-operations-absolute}. The x-axis (linear) on both figures is the desired traffic coverage. In the occupancy calculation, we assumed that the average packet interarrival time is the same for all flows. The presented results of analytical calculations are in line with selected values obtained in simulations, which confirms their correctness.

\begin{figure*}[!htb]
\centering
\includegraphics[scale=0.46,right]{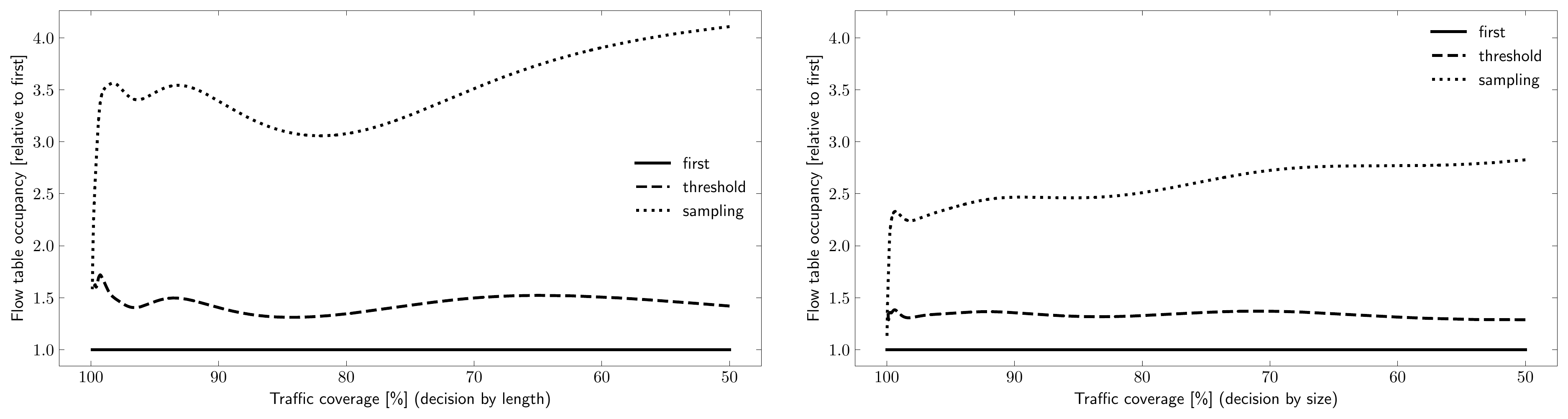}
\caption{Analytically calculated flow table occupancy relative to the \emph{first} algorithm in function of traffic coverage.}
\label{fig-occupancy-first}
\end{figure*}

\begin{figure*}[!htb]
\centering
\includegraphics[scale=0.46,right]{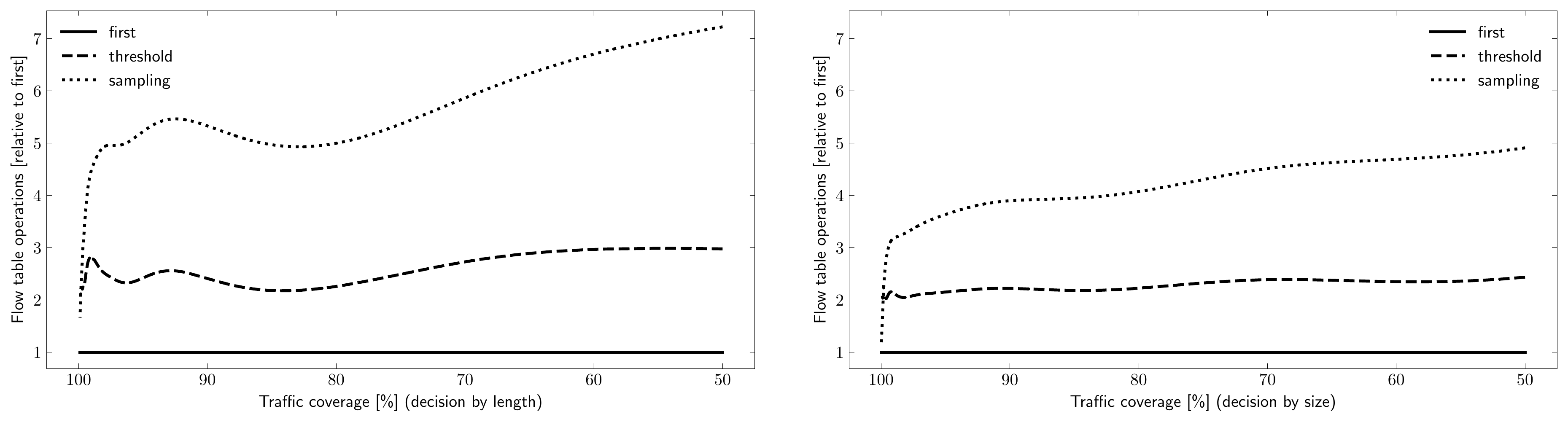}
\caption{Analytically calculated flow table operations number relative to the \emph{first} algorithm in function of traffic coverage.}
\label{fig-operations-first}
\end{figure*}

\begin{table*}[!ht]
\caption{Classes of algorithms for which each analyzed approach provides boundaries}
\scriptsize
\begin{center}
\begin{tabular}{@{}lp{4cm}p{4cm}p{4cm}@{}}
\toprule
& \emph{first} & \emph{threshold} & \emph{sampling} \\
\midrule
Upper boundary & All algorithms, machine learning algorithms based on header classification & Exact counters, inexact counters, Bloom filters, sketches & \\
\midrule
Lower boundary & & & All algorithms classifying flows not with the first, but on some subsequent packet \\
\bottomrule
\end{tabular}
\label{tab-classes}
\end{center}
\end{table*}

It can be seen that the \emph{first} algorithm achieves the best performance. For any target traffic coverage, it gives the largest reduction, both in flow table occupancy and operations number. Additionally, in the case of the \emph{first} algorithm, operations number and occupancy reduction factors are the same. While its implementation is impossible in practice, the results are still valuable as they provide an upper boundary for all flow table usage reduction algorithms based on elephant/mice classification. Because of that, on \autoref{fig-occupancy-first} and \autoref{fig-operations-first}, we also show the analytically calculated performance of all algorithms relative to it.

In the case of \emph{threshold} algorithm, the average reduction of flow table occupancy is always higher than the reduction of the number of flow entry operations. This is expected as flows are added not with their first packet but with some subsequent packet, so they occupy the flow table for only a fraction of their lifetime. Unfortunately, the same applies to traffic coverage. To achieve similar traffic coverage, a lower threshold value has to be used than in the \emph{first} approach, which also results in lower table occupancy reduction. Occupancy in the case of \emph{threshold} algorithm is approximately 1.5 times higher than with the \emph{first} and the number of flow entries operations is 2-3 times higher. Its usability is limited in OpenFlow switches, but low-overhead implementation may be possible in other dataplane technologies (such as P4). Apart from that, it provides an upper boundary for all algorithms based on some kind of inexact counting, for example, Bloom filters or sketches.

In the case of \emph{sampling} algorithm, similarly as in \emph{threshold}, the average reduction of flow table size is higher than the reduction of the number of flow entry operations, for the same reason. It can be also seen that, in general, decreasing sampling probability results in an exponential increase in the flow table occupancy reduction, but only in a linear decrease in traffic coverage. In the case of \emph{sampling} algorithm, occupancy is 2-4 times higher and operations number is 3-7 times higher than in the \emph{first}. However, the numbers are still high. For example, with a target of 80\% traffic coverage, which we believe is a fair target in TE case, it can reduce the number of flow table entries by a factor of 225 (packet sampling) or 401 (size-based non-uniform sampling). Thus, the performance of \emph{sampling} stays within the same order of magnitude as the performance of \emph{first} and \emph{threshold} algorithms. Unlike them, it is trivial to implement, has negligible computational overhead, and does not require any memory.

Finally, all algorithms achieve better performance when \emph{size} is used as a threshold/sampling base. This can be attributed to the fact that longer flows have a larger average packet size.

\section{Further research}
\label{further}

In this paper, we use flow sizes to calculate traffic coverage. However, it would be more beneficial to maximize the coverage of flow rates. It has been shown that there is a correlation between flow sizes and rates \cite{zhang2002characteristics}. Therefore, the idea that high-rate flows could be identified using flow sizes is theoretically correct \cite{mori2004identifying}. Nevertheless, it may be interesting to directly use flow rates for traffic coverage calculations.

Similarly, we assumed that packet interarrival time is the same for all flows and does not depend on flow length or size, in other words, that the flow duration is proportional to flow length/size. Using real flow durations to calculate occupancy would give more accurate results. However, the calculation of flow rate and duration distributions requires accurate timestamps. Hardware NetFlow agents usually cannot assign accurate timestamps to generated flow records \cite{hofstede2014flow}. Therefore, packet traces instead of flow records should be analyzed in order to obtain accurate flow rate and duration distributions. Unfortunately, sufficiently long packet traces are not publicly available (CAIDA traces, as we already stated, are truncated to one hour).


In our research, we followed the NetFlow default 15 seconds inactive timeout. It will be interesting to perform similar evaluations for flows with other timeouts, especially subsecond (so-called \emph{flowlets}). Flowlet-based traffic engineering is an interesting concept since the number of simultaneous flowlets is several orders of magnitude lower than the number of simultaneous flows. However, we also expect that gains from elephant-mice flow differentiation will be lower for flowlets.

Another interesting research direction is the usage of elephant-based traffic engineering in distributed systems. In centralized approaches, all switches send detected elephants to the central controller, which installs flow-specific paths on all switches at once. In distributed systems, the installation of elephant flow-specific entry would have to be coordinated between all switches without the usage of a central controller.

Implementation aspects are also very important. The exact counting of packets of all candidate flows is resource-intensive in OpenFlow switches. Novel dataplane technologies, like P4 or eBPF, can allow the implementation of low-overhead per-flow counters \cite{turboflow}. An interesting alternative is the inexact counting of significant queue contributors based on count-min sketches, which was proven to run on P4 switch at line rate and identify elephants with high accuracy and low latency \cite{10.1145/3359989.3365408}. High performance and low latency out-of-band sampling mechanisms, like \cite{rasley2014planck}, are also an interesting research topic.

\section{Conclusions}
\label{conclusions}

The contribution of this paper is fourfold. First, it examines elephant-based flow table usage reduction approaches from the traffic engineering point of view. The idea, although already appeared as a part of some TE systems, was not analyzed numerically in isolation. It remained unknown what extent of flow table occupancy and operations number reduction can be achieved, how to select thresholds or sampling rates to cover the desired fraction of traffic.

Secondly, the presented results of the \emph{first} and \emph{threshold} algorithms provide upper boundaries for whole classes of flow table usage reduction algorithms. The \autoref{tab-classes} presents boundaries provided by each of the analyzed algorithms. Such boundaries were not previously available.

Thirdly, we discovered the surprisingly good performance of the \emph{sampling} algorithm. It introduces a negligible overhead to the packet processing pipeline and does not require any memory (unlike counters or Bloom filters), yet when applied to realistic traffic, it can reduce the number of flow entries by a factor of 400, while still maintaining 80\% of the traffic covered by individual flow entries. Thus, it can provide a lower bound for the performance of other algorithms and in cases when flow rerouting is acceptable, it can eliminate the need for more sophisticated solutions.

Finally, the key aspect of our analysis is the usage of accurate and reproducible flow length and size distribution mixtures. The accuracy of used distributions has a crucial impact on results. We acknowledge that various networks can have different distributions; therefore, we provide an open-source software package allowing both the replication of our experiments and the performing of similar evaluations for other algorithms or flow distributions \cite{github-flow-models}.

The elephant detection for forwarding purposes is becoming currently an attractive research topic (e.g. ML-based detection in P4). However, to analyze more sophisticated solutions, baselines and theoretical boundaries need to be established first. This is the goal of this paper. It sets the baseline for analysis of the performance of flow table occupancy reduction algorithms of various classes and provide methodology, traffic model and software which can be used by other researchers.

\section*{Acknowledgment}
\noindent
The research was carried out with the support of the project "Intelligent management of traffic in multi-layer Software-Defined Networks" founded by the Polish National Science Centre under project no. 2017/25/B/ST6/02186.

\onecolumn
\appendix
\section{Simulation results}
\label{appendix}

\noindent The tables below present packet-level simulation results for all analyzed algorithms. They can be also used as reference for researchers and network operators on how to set threshold values or sampling probabilities to achieve a desired traffic coverage or flow table usage reduction.

\begin{table*}[!h]
\caption{Simulation results (decision by length)}
\scriptsize
\begin{center}
\begin{tabular}{@{}lrrrrrrlrrr@{}}
\cmidrule[0.75pt](r){1-7} \cmidrule[0.75pt](l){8-11}
\multirow{4.4}{1.28cm}{\textbf{Threshold (packets)}} & \multicolumn{3}{c}{\textbf{First algorithm}} & \multicolumn{3}{c}{\textbf{Threshold algorithm}} & \multirow{4.4}{1.2cm}{\textbf{Sampling prob.}} & \multicolumn{3}{c}{\textbf{Sampling algorithm}} \\
\cmidrule(lr){2-4} \cmidrule(lr){5-7} \cmidrule(l){9-11}
& Traffic &       Operations &       Occupancy & Traffic &       Operations &       Occupancy & & Traffic &       Operations &       Occupancy \\
& coverage & reduction & reduction & coverage & reduction & reduction & & coverage & reduction & reduction\\
& (\%) & (x) & (x) & (\%) & (x) & (x) & & (\%) & (x) & (x)\\
\cmidrule[0.5pt](r){1-7} \cmidrule[0.5pt](l){8-11}
1       &       99.89 &            1.92 &           1.92 &       99.71 &            1.92 &           2.60 & 1.00 &      100.00 &            1.00 &           1.00 \\
2       &       99.82 &            2.88 &           2.88 &       99.52 &            2.88 &           4.06 & 5.00e-01 &       99.77 &            1.41 &           1.54 \\
4       &       99.74 &            3.89 &           3.89 &       99.23 &            3.89 &           6.16 & 2.50e-01 &       99.47 &            2.04 &           2.41 \\
8       &       99.56 &            5.99 &           5.99 &       98.77 &            5.99 &          10.28 & 1.25e-01 &       99.04 &            3.00 &           3.81 \\
16      &       99.22 &           10.40 &          10.40 &       98.10 &           10.40 &          17.71 & 6.25e-02 &       98.43 &            4.53 &           6.09 \\
32      &       98.75 &           17.32 &          17.32 &       97.16 &           17.32 &          29.15 & 3.12e-02 &       97.61 &            6.93 &           9.74 \\
64      &       97.99 &           28.33 &          28.33 &       95.87 &           28.33 &          46.66 & 1.56e-02 &       96.46 &           10.83 &          15.78 \\
128     &       96.99 &           44.05 &          44.05 &       94.16 &           44.05 &          73.62 & 7.81e-03 &       94.97 &           16.95 &          25.42 \\
256     &       95.65 &           69.57 &          69.57 &       91.88 &           69.57 &         119.93 & 3.90e-03 &       92.96 &           26.88 &          41.30 \\
512     &       93.79 &          115.98 &         115.98 &       88.88 &          115.98 &         198.05 & 1.95e-03 &       90.37 &           42.21 &          66.07 \\
1 024    &       91.44 &          191.38 &         191.38 &       84.96 &          191.38 &         318.15 & 9.76e-04 &       86.93 &           67.57 &         107.39 \\
2 048    &       88.45 &          300.49 &         300.49 &       79.73 &          300.49 &         503.95 & 4.88e-04 &       82.52 &          105.88 &         170.58 \\
4 096    &       84.16 &          469.59 &         469.59 &       72.77 &          469.59 &         827.40 & 2.44e-04 &       76.41 &          169.31 &         276.96 \\
8 192    &       77.78 &          775.64 &         775.64 &       64.01 &          775.64 &        1462.54 & 1.22e-04 &       69.26 &          271.58 &         453.25 \\
16 384   &       69.37 &         1399.51 &        1399.51 &       53.83 &         1399.51 &        2834.49 & 6.10e-05 &       61.21 &          431.17 &         735.66 \\
32 768   &       59.27 &         2794.15 &        2794.15 &       42.60 &         2794.15 &        6069.15 & 3.05e-05 &       50.30 &          727.99 &        1271.26 \\
65 536   &       47.29 &         6201.41 &        6201.41 &       31.09 &         6201.41 &       14399.51 & 1.52e-05 &       40.64 &         1229.39 &        2197.34 \\
131 072  &       34.27 &        15345.62 &       15345.62 &       20.65 &        15345.62 &       37977.86 & 7.62e-06 &       30.39 &         2283.27 &        4198.14 \\
262 144  &       22.41 &        42262.61 &       42262.61 &       12.47 &        42262.61 &      111279.85 & 3.81e-06 &       19.61 &         4994.85 &        9425.16 \\
524 288  &       13.26 &       130950.45 &      130950.45 &        6.84 &       130950.45 &      367074.49 & 1.90e-06 &       14.21 &         8402.36 &       16061.45 \\
1 048 576 &        7.09 &       456577.43 &      456577.43 &        3.37 &       456577.43 &     1365306.53 & 9.53e-07 &        9.95 &        14669.12 &       28322.65 \\
2 097 152 &        3.37 &      1799949.44 &     1799949.44 &        1.49 &      1799949.44 &     5604593.17 & 4.76e-07 &        6.21 &        27264.93 &       53215.32 \\
\cmidrule[0.75pt](r){1-7} \cmidrule[0.75pt](l){8-11}
\end{tabular}
\label{tab-length}
\end{center}
\end{table*}

\begin{table*}[!h]
\caption{Simulation results (decision by size)}
\scriptsize
\begin{center}
\begin{tabular}{@{}lrrrrrrlrrr@{}}
\cmidrule[0.75pt](r){1-7} \cmidrule[0.75pt](l){8-11}
\multirow{4.4}{1.28cm}{\textbf{Threshold (bytes)}} & \multicolumn{3}{c}{\textbf{First algorithm}} & \multicolumn{3}{c}{\textbf{Threshold algorithm}} & \multirow{4.4}{1.2cm}{\textbf{Sampling prob.}} & \multicolumn{3}{c}{\textbf{Sampling algorithm}} \\
\cmidrule(lr){2-4} \cmidrule(lr){5-7} \cmidrule(l){9-11}
& Traffic &       Operations &       Occupancy & Traffic &       Operations &       Occupancy & & Traffic &       Operations &       Occupancy \\
& coverage & reduction & reduction & coverage & reduction & reduction & & coverage & reduction & reduction\\
& (\%) & (x) & (x) & (\%) & (x) & (x) & & (\%) & (x) & (x)\\
\cmidrule[0.5pt](r){1-7} \cmidrule[0.5pt](l){8-11}
64         &      100.00 &            1.04 &           1.04 &       99.90 &            1.04 &           1.58 & 1.00 &      100.00 &            1.00 &           1.00 \\
128        &       99.95 &            1.53 &           1.53 &       99.83 &            1.53 &           2.45 & 5.00e-01 &       99.98 &            1.12 &           1.14 \\
256        &       99.89 &            2.34 &           2.34 &       99.73 &            2.34 &           3.58 & 2.50e-01 &       99.90 &            1.49 &           1.61 \\
512        &       99.82 &            3.43 &           3.43 &       99.59 &            3.43 &           5.12 & 1.25e-01 &       99.76 &            2.06 &           2.38 \\
1 024       &       99.73 &            4.76 &           4.76 &       99.41 &            4.76 &           7.23 & 6.25e-02 &       99.57 &            2.86 &           3.51 \\
2 048       &       99.60 &            6.74 &           6.74 &       99.14 &            6.74 &          10.60 & 3.12e-02 &       99.32 &            4.01 &           5.18 \\
4 096       &       99.38 &           10.03 &          10.03 &       98.77 &           10.03 &          15.83 & 1.56e-02 &       98.98 &            5.67 &           7.66 \\
8 192       &       99.10 &           15.02 &          15.02 &       98.28 &           15.02 &          23.52 & 7.81e-03 &       98.54 &            8.08 &          11.32 \\
16 384      &       98.73 &           22.18 &          22.18 &       97.62 &           22.18 &          34.87 & 3.90e-03 &       97.94 &           11.69 &          16.85 \\
32 768      &       98.22 &           32.86 &          32.86 &       96.73 &           32.86 &          52.09 & 1.95e-03 &       97.14 &           16.99 &          25.07 \\
65 536      &       97.52 &           49.23 &          49.23 &       95.53 &           49.23 &          78.37 & 9.76e-04 &       96.08 &           24.87 &          37.35 \\
131 072     &       96.58 &           74.00 &          74.00 &       93.93 &           74.00 &         118.38 & 4.88e-04 &       94.70 &           36.29 &          55.38 \\
262 144     &       95.32 &          111.45 &         111.45 &       91.80 &          111.45 &         180.52 & 2.44e-04 &       92.84 &           53.78 &          83.23 \\
524 288     &       93.60 &          170.39 &         170.39 &       89.00 &          170.39 &         279.56 & 1.22e-04 &       90.43 &           79.91 &         125.35 \\
1 048 576    &       91.29 &          265.01 &         265.01 &       85.37 &          265.01 &         437.69 & 6.10e-05 &       87.36 &          119.62 &         189.44 \\
2 097 152    &       88.29 &          414.00 &         414.00 &       80.71 &          414.00 &         690.84 & 3.05e-05 &       83.22 &          182.75 &         292.31 \\
4 194 304    &       84.41 &          650.85 &         650.85 &       74.77 &          650.85 &        1112.19 & 1.52e-05 &       77.99 &          275.65 &         448.04 \\
8 388 608    &       79.21 &         1053.94 &        1053.94 &       67.30 &         1053.94 &        1847.74 & 7.62e-06 &       71.80 &          415.82 &         685.60 \\
16 777 216   &       72.51 &         1759.10 &        1759.10 &       58.24 &         1759.10 &        3152.64 & 3.81e-06 &       64.19 &          659.01 &        1096.09 \\
33 554 432   &       64.46 &         2945.28 &        2945.28 &       47.41 &         2945.28 &        5653.28 & 1.90e-06 &       53.81 &         1052.41 &        1793.71 \\
67 108 864   &       53.75 &         5289.09 &        5289.09 &       34.77 &         5289.09 &       11787.26 & 9.53e-07 &       43.64 &         1804.65 &        3155.34 \\
134 217 728  &       38.69 &        12142.54 &       12142.54 &       22.15 &        12142.54 &       31499.08 & 4.76e-07 &       32.17 &         3176.87 &        5777.40 \\
268 435 456  &       23.50 &        36943.89 &       36943.89 &       12.63 &        36943.89 &      102107.04 & 2.38e-07 &       22.36 &         5640.73 &       10524.92 \\
536 870 912  &       13.20 &       124827.05 &      124827.05 &        6.77 &       124827.05 &      350449.76 & 1.19e-07 &       14.24 &         9935.61 &       18946.22 \\
1 073 741 824 &        7.02 &       434921.71 &      434921.71 &        3.33 &       434921.71 &     1273242.43 & 5.96e-08 &       10.50 &        17532.00 &       33966.28 \\
\cmidrule[0.75pt](r){1-7} \cmidrule[0.75pt](l){8-11}
\end{tabular}
\label{tab-size}
\end{center}
\end{table*}

\twocolumn

\bibliographystyle{IEEEtran}
\bibliography{./bib}

\end{document}